\documentclass[10pt,journal,compsoc]{IEEEtran}

\ifCLASSOPTIONcompsoc
  \usepackage[nocompress]{cite}
\else
  \usepackage{cite}
\fi

\usepackage{comment}

\ifCLASSINFOpdf
   \usepackage[pdftex]{graphicx}
   \DeclareGraphicsExtensions{.pdf,.jpeg,.png}
\else
   \usepackage[dvips]{graphicx}
   \DeclareGraphicsExtensions{.eps}
\fi

\ifCLASSOPTIONcompsoc
  \usepackage[caption=false,font=footnotesize,labelfont=sf,textfont=sf]{subfig}
\else
  \usepackage[caption=false,font=footnotesize]{subfig}
\fi

\usepackage{url}
\usepackage{amssymb}
\usepackage{amsmath}
\usepackage{algorithm,algpseudocode}
\usepackage{multirow}
\usepackage[table]{xcolor}
\algnewcommand\algorithmicforeach{\textbf{for each}}
\algdef{S}[FOR]{ForEach}[1]{\algorithmicforeach\ #1\ \algorithmicdo}

\usepackage{tcolorbox}
\tcbset{
  colback=gray!10,
  colframe=gray!60,
  boxrule=0.6pt,
  arc=2pt,
  left=6pt,right=6pt,top=4pt,bottom=4pt
}

\makeatletter
\def\@IEEENORMtitlevspace{0.1\baselineskip} 
\def\@IEEEMINtitlevspace{0.1\baselineskip}
\makeatother

\begin{document}
\title{Predicting Student Actions in a Procedural Training Environment}
%
%
%
%

\author{Diego~Riofr\'{i}o-Luzcando,
		Jaime~Ram\'{i}rez
        and~Marta~Berrocal-Lobo
\IEEEcompsocitemizethanks{
\IEEEcompsocthanksitem D. Riofr\'{i}o-Luzcando and J. Ram\'{i}rez are with ETS de Ingenieros Inform\'{a}ticos, UPM, Spain.\protect\\
E-mail: \{driofrio, jramirez\}@fi.upm.es
\IEEEcompsocthanksitem Marta Berrocal-Lobo is with E.T.S.I. Forestal y del Medio Natural, UPM, Spain, and Center for Plant Biotechnology and Genomics, UPM-INIA, Spain.\protect\\
E-mail: m.berrocal@upm.es
}
}

\IEEEaftertitletext{%
\begin{tcolorbox}
\small
\textbf{Author Accepted Manuscript (AAM).} This is the accepted version of the manuscript.\\
Please cite as: Riofr\'{i}o-Luzcando, D.; Ram\'{i}rez, J.; Berrocal-Lobo, M. (2017).
\emph{Predicting Student Actions in a Procedural Training Environment}. \emph{IEEE Transactions on Learning Technologies}.\\
Final published version available at: \texttt{https://doi.org/10.1109/TLT.2017.2658569}.\\[2pt]
\textcopyright\ 2017 IEEE. Personal use of this material is permitted. Permission from IEEE must be obtained for all other uses, in any current or future media,
including reprinting/republishing this material for advertising or promotional purposes, creating new collective works, for resale or redistribution to servers or lists,
or reuse of any copyrighted component of this work in other works.
\end{tcolorbox}
\vspace{2.5em}
}

\IEEEtitleabstractindextext{%
\begin{abstract}
Data mining is known to have a potential for predicting user performance. However, there are few studies that explore its potential for predicting student behavior in a procedural training environment. This paper presents a collective student model, which is built from past student logs. These logs are firstly grouped into clusters. Then an extended automaton is created for each cluster based on the sequences of events found in the cluster logs. The main objective of this model is to predict the actions of new students for improving the tutoring feedback provided by an intelligent tutoring system. The proposed model has been validated using student logs collected in a 3D virtual laboratory for teaching biotechnology. As a result of this validation, we concluded that the model can provide reasonably good predictions and can support tutoring feedback that is better adapted to each student type.
\end{abstract}

}

\maketitle

\IEEEdisplaynontitleabstractindextext
\IEEEpeerreviewmaketitle

\IEEEraisesectionheading{\section{Introduction}\label{sec:introduction}}
\IEEEPARstart{E}{ducational} data mining has already achieved promising results, for example, with regard to the analysis of student performance or the prediction of student grades, especially in the field of web e-learning \cite{Romero2007,Romero2010}. However, there is hardly any research in the literature that has integrated data mining techniques into intelligent tutoring systems (ITSs) \cite{Baker2014}, for example, to provide customized tutoring for each student.

This paper presents a collective student model that has been designed to anticipate the actions that students are likely to take while completing a practical assignment in an educational environment for procedural training. This model is created from activity records or logs collected from students with a similar background that previously completed the same practical assignment. As we will see later, an ITS equipped with this collective student model can use hints to stop students from making certain errors or from floundering with the practical assignment.

It is sometimes a good idea to let students make mistakes from which they learn. In other cases, however, it is better to give students the minimum amount of support that they need to progress independently towards problem solving and overcome their zones of proximal development \cite{Vygotsky1978}. In this way, each student learns not from his or her mistakes but with a little bit of help. If necessary, the tutor gradually increases the level of support or scaffolding every time the student makes a mistake or gradually reduces the amount of help provided when the student makes progress \cite{Olneygift2014,Holdengift2014}. Another reason for helping students not to make mistakes is to prevent student frustration when they fail too often.

The proposed collective student model consists of several clusters of students (Section \ref{sec:collect_student_model}), each of which contains an extended automaton (Section \ref{sec:automaton}). This automaton is a directed graph adapted for our purposes. As explained later, these clusters will help to provide automatic tutoring adapted to each student type. In order to confirm this claim, we validated the model using student logs collected in a 3D virtual laboratory for teaching biotechnology. This validation had two main goals: i) verify that the prediction error is acceptable for tutoring purposes; and ii) check whether clustering methods can classify students into groups that require different tutoring feedback. As we will see later, although students had a lot of freedom of action in this virtual laboratory, the model was reasonably reliable at predicting student actions and provided a useful classification of students into clusters according to their performance.

The structure of the remainder of the paper is as follows. Section \ref{sec:related_works} shows relevant works in the field of educational data mining. Section \ref{sec:architecture_proposed} describes the proposed ITS architecture, which would be able to leverage the collective student model detailed later in Section \ref{sec:collect_student_model}. Section \ref{sec:model_val} reports model validation detailing the method followed in this study and discussing its results. Finally, Section \ref{sec:conclusion} outlines the conclusions of this research and some future work.

\section{Related Work}\label{sec:related_works}

The related work is divided into two sections. Section \ref{ssec:EDM} briefly presents the main goals of educational data mining and mentions some of the key results with respect to web based e-learning systems. Section \ref{ssec:ITS} focuses on systems for procedural training equipped with ITSs, whose student logs have been analyzed by means of data mining.

\subsection{Educational Data Mining (EDM)}\label{ssec:EDM}
EDM tries to use data sourced from the repositories of different types of learning environments to better understand learners and learning \cite{Romero2010}. Some general applications of EDM are \cite{Romero2010a}: communicate student activities and usage of online courses to educational stakeholders; help with course maintenance and improvement by analyzing usage data; analyze how well the domain is structured by student performance prediction; generate recommendations for students; predict student grades and learning outcomes; and model students. Given the scope of our research, the literature review will focus on the last three EDM applications.

Some researchers \cite{Perera2009,Tang2005, Godoy2010} use data mining to provide hints, feedback or recommendations about which content is best for each student. Some of these use an ITS \cite{Fournier-Viger2013}. The most frequently used data mining techniques for this purpose are association, sequencing, classification and clustering.

Other researchers \cite{Dekker2009, Romero2013a, Lara2014} try to predict different kinds of student learning outcomes such as final grades or dropouts. The most frequent data mining techniques used in this group are association, classification and clustering.

Student modeling \cite{Antunes2008, Hershkovitz2009, Arroyo2009, Barnes2010, Mavrikis2010, Porayska-Pomsta2013} has several applications such as the detection of student behavior or learning problems. This group most frequently uses the same data mining techniques as above, plus statistical analyses, Bayes networks, psychometric models and reinforcement learning.

One noteworthy paper in the last group processes Moodle logs to discover a specific student behavior model \cite{Bogarin2014}. They divide these logs into student groups with similar characteristics using a clustering method and then apply process mining to each cluster to create a model (represented by a directed acyclic graph) that shows the most frequent sequences of student actions. An interesting conclusion of this paper, which is relevant for our research, is that graphs, models or visual representations are easier to comprehend. Teachers and students find this summarized information more accessible. Therefore, this information could be very useful for monitoring the learning process and providing feedback.

As we find from the works referenced in this section, most research in EDM has focused on studying data or logs registered by web e-learning systems, like Moodle or MOOCs, or data collected from student curricula \cite{Romero2007,Romero2010,Pena-Ayala2014,Sukhija2015}.

To the best of our knowledge, there is no any other proposal of predictive model in the literature to support procedural training environments that relies on data mining. Hence, next section will focus on a closed related area where we have actually been able to find some interesting contributions, procedural training environments equipped with ITSs, whose student logs have been processed through data mining.

\subsection{Intelligent Tutoring Systems with EDM in Procedural Training}\label{ssec:ITS}
Two well-known ITSs that employ EDM are Assistment \cite{Razzaq2005} and Cognitive Tutor Algebra I \cite{Ritter2007,Ritter2015}. Both are web tools that guide students through the process of solving math exercises. There are several data mining studies using data collected from these two environments, but they do not report whether or not the results of these studies have been used to improve the tutoring services. For example, data from Assistment were used to create a model to predict when a student is about to ask for a hint \cite{vicente2015}. EDM is applied in Cognitive Tutor Algebra I to create a model that detects student attitudes/feelings such as engagement, concentration, confusion, frustration, and boredom solely from student interactions within the tutor \cite{Baker2012}.

To the best of our knowledge, there is only one learning environment for procedural training equipped with an ITS that relies on EDM. It is called CanadarmTutor \cite{Fournier-Viger2013}. It simulates the Canadarm2 robotic arm used in the International Space Station. This ITS provides assistance to users on how to perform a correct sequence of arm operations to reach a goal. To do this, it integrates a cognitive model to assess skills and spatial reasoning, and an expert system that automatically generates solution paths. Because of the ill-defined characteristics of the problem-solving procedure, the ITS uses data mining techniques to extract a partial task model from past user solutions. Using this model the ITS can recognize the learner plan and provides assistance based on the prediction of the user's next action.

Despite the research that has already been conducted in this area, the community is missing more generally applicable results \cite{Romero2013}, for example, predictive models that can be used in more than one different context. There is also a remarkable shortage of intelligent educational systems that take advantage of models developed by EDM \cite{Baker2014}.

The research presented in this paper represents a step forward towards the development of an ITS that leverages a collective student model computed by means of EDM to offer better tutoring feedback. Moreover, this model is intended for procedural training in learning environments and is domain independent.

\section{ITS Architecture Proposal}\label{sec:architecture_proposed}
\begin{figure}[htb]
 \centering
 \includegraphics[width=\columnwidth]{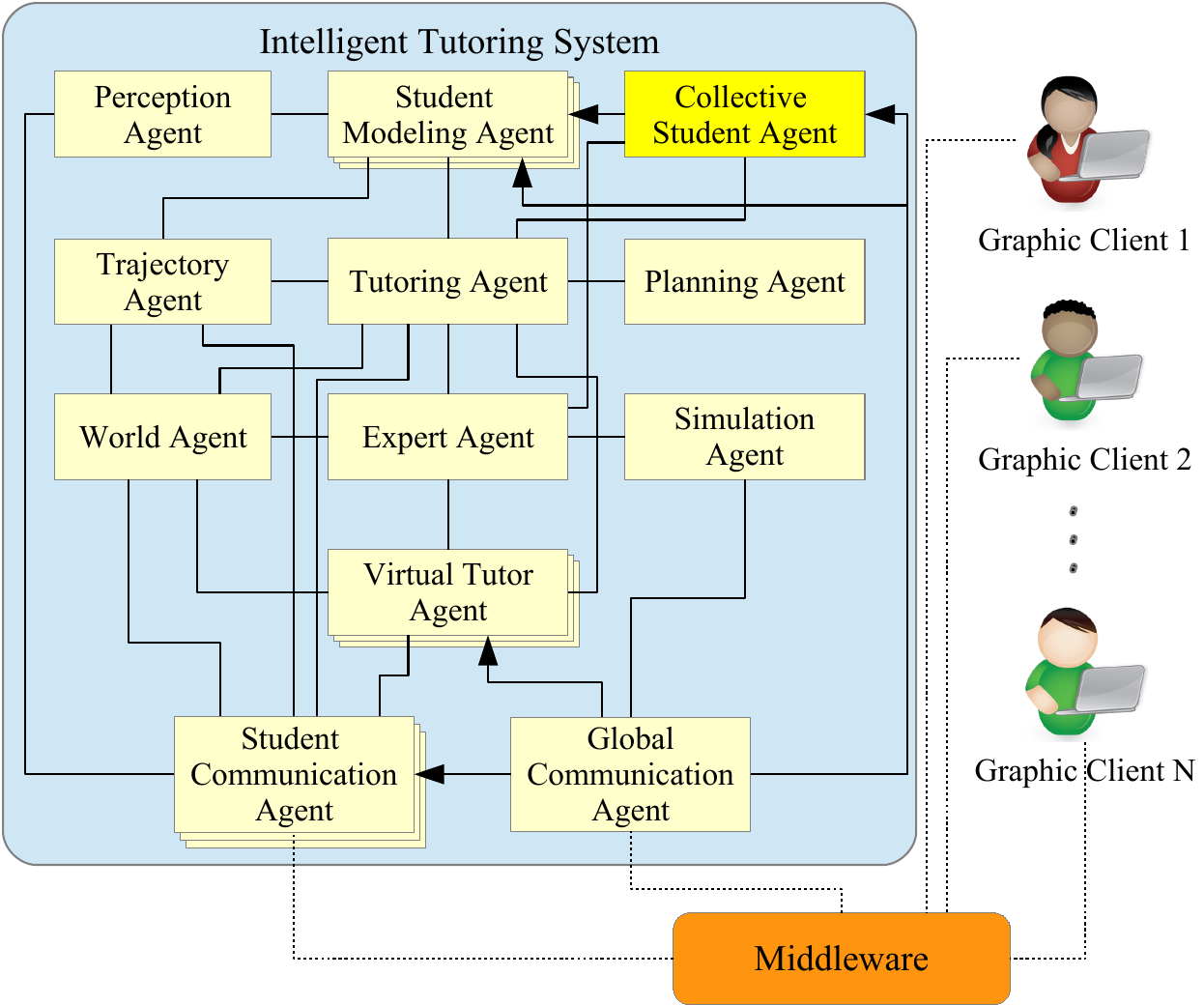}
 \caption{MAEVIF architecture with the new agent}
 \label{fig:architecture}
\end{figure}

In order to leverage the presented collective student model, we propose an extension of a previous ITS architecture, the MAEVIF architecture \cite{DeAntonio2005, Imbert2007, Clemente2011}, which is depicted in Figure \ref{fig:architecture}. MAEVIF is a multi-agent architecture that is an adaptation of the classic ITS architecture for learning environments specialized in training. Within the extension of MAEVIF, this model encompasses a new agent, called Collective Student Agent.

Of the MAEVIF agents, let us focus on the Student Modeling Agent \cite{Clemente2011a, Clemente2011, Clemente2014}. The Student Modeling Agent contains an adaptive, extensible and reusable student model that infers the student knowledge state using a pedagogic-cognitive diagnosis with non-monotonic reasoning abilities. The purpose of this agent is to discover each student's learning status, that is, what he or she does or does not know about the subject. This can serve as support for the personalized automatic tutoring of each student. In this way, if the student model contains enough information on a particular student, it will provide good predictions of his/her behavior. For example, if the student model knows that a student recently performed a task correctly, it is very likely that this student will perform the same (or a very similar) task correctly again.

One disadvantage of this student model is, however, that, if queried about the attainment level of a particular learning objective, it will need a lot of background information about the student with respect to that learning objective in order to give a reliable enough response, and this information will often not be available. For example, this may be the case if it is the student's first attempt at an exercise. This may constitute a problem when the tutoring agent needs to predict the student's next actions, because if the student modeling agent is not confident enough that the student knows which actions to take next, it will not be able to provide a good prediction.

As this paper shows, if the student model does not possess enough information on a particular student, the collective student model will be a reasonably good alternative. The collective student model comprises summarized data on past student action events that are used to predict the actions that a student under supervision is most likely to take next. The premise for creating this model is that the behavior of past students doing a practical assignment should be similar to current students with the same training completing the same practical assignment.

\section{Description of the Collective Student Model}\label{sec:collect_student_model}
This model is created using historical data from past students and is continually refined with the actions from students under supervision. Our model can be considered as the result of the models/patterns discovery phase of the knowledge discovery in databases process adapted to EDM as formulated by Romero and Ventura \cite{Romero2013}.

The idea of building this model was inspired by our experiences evaluating the 3D virtual lab for biotechnology \cite{Rico2012}. During model design, we observed the behavior of students in the virtual world following the ethnographic method. Subsequently, as recommended by Mostow et al. \cite{Mostow2006}, the student logs were analyzed by hand to identify interesting phenomena. One of the conclusions drawn from this analysis was that students tend to fall into different groups depending on their performance in the practical assignment. In addition, it was clear that different groups required different levels of help to complete the practical assignment. This idea led us to apply clustering techniques to student logs.

\begin{figure}[!htb]
 \centering
 \includegraphics[width=\columnwidth]{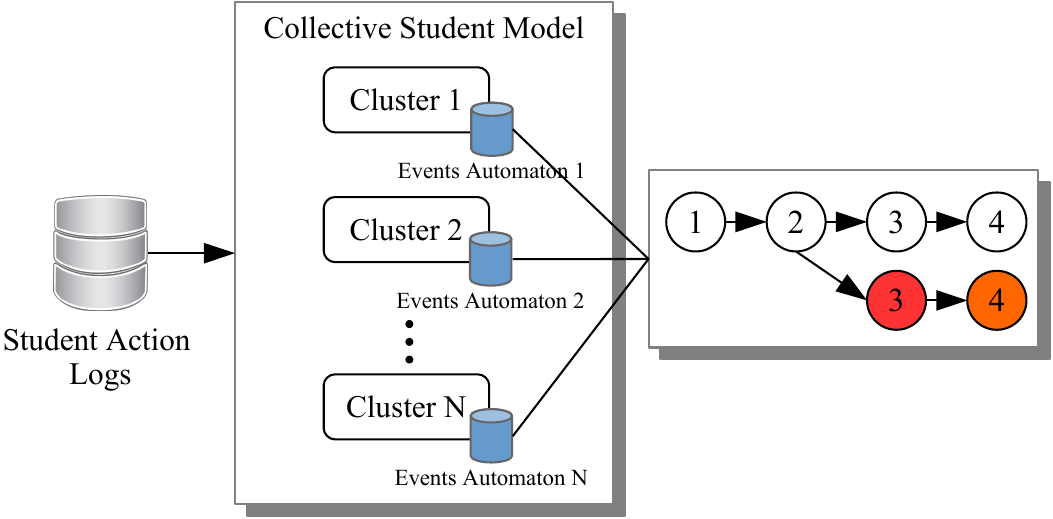}
 \caption{Collective Student Model}
 \label{fig:collect_model}
\end{figure}

Besides, sequence models can be viewed in data mining as graph-based models \cite{Mannila2000}, and they could be described as directed graphs for discovering frequent events \cite{Mannila1997}. Hence, the proposed model (see Figure \ref{fig:collect_model}) consists of several clusters of students, each of which contains an extended automaton (detailed in Section \ref{sec:automaton}), which is a directed graph fit for our purposes. As discussed later, these clusters will help to provide automatic tutoring adapted to each student type.

Model creation shares the same main phases as the process proposed by Bogar\'in et al. \cite{Bogarin2014}. It is executed when the tutor is launched. To create the model, it is necessary to access student log events stored in the student model ontology \cite{Clemente2011a, Clemente2011}. Firstly, student logs are clustered based on any of the clustering methods detailed in Section \ref{sec:clumet}. Then, for each cluster, an automaton is built from the student logs of this cluster. Next, the model is updated with each new student action at training time. In this way, the model can adapt to the students under supervision better and therefore deal with differences of behavior between current students and past students.

\subsection{Extended Automaton Definition}\label{sec:automaton}
\begin{figure*}[!h]
 \centering
 \includegraphics[width=\textwidth]{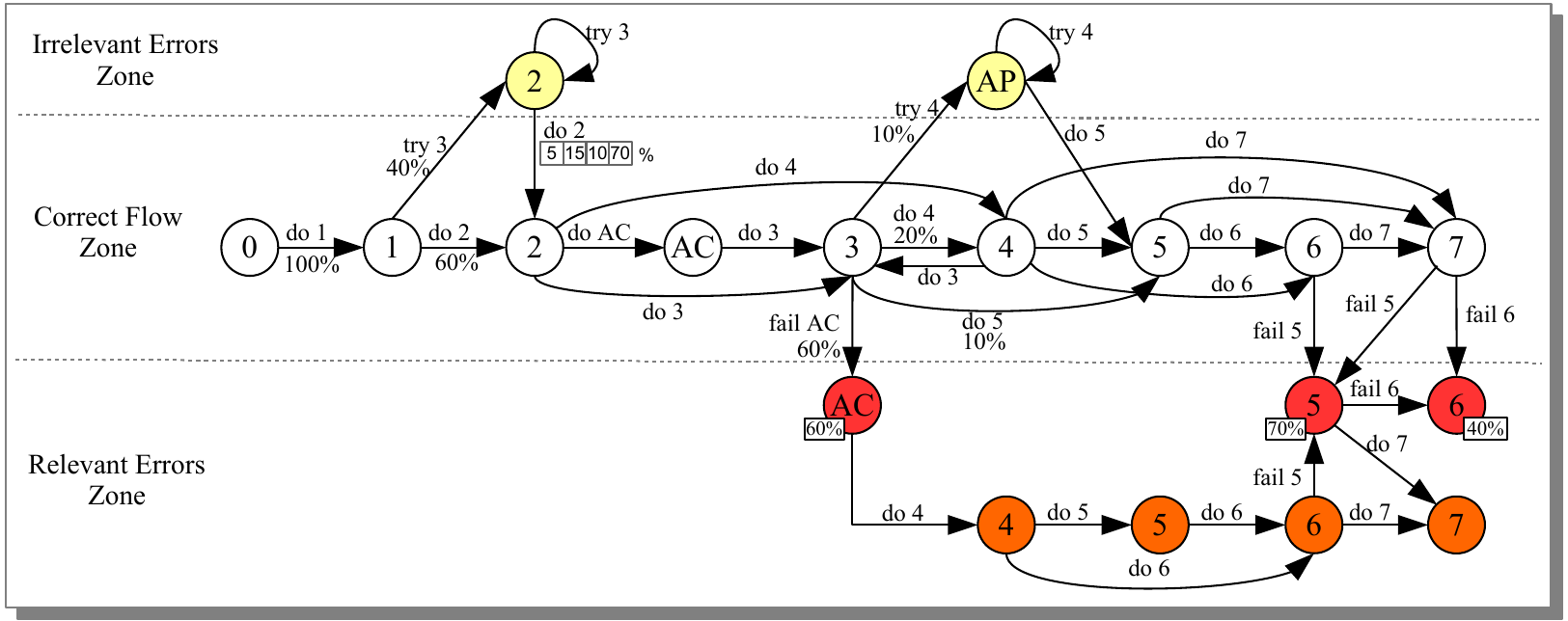}
 \caption{Example automaton}
 \label{fig:example_automaton}
\end{figure*}

States are represented by circles and transitions by arrows as shown in Figure \ref{fig:example_automaton}. Furthermore, states are grouped into three zones: Correct Flow Zone, Irrelevant Errors Zone and Relevant Errors Zone.

Transitions denote events generated by students throughout an exercise, such as actions or attempted actions that past students have performed so far and new students may repeat in the future. Within the automaton, an event represents one of the following situations:

\begin{itemize}
\item a valid action for an exercise (do X event in Figure \ref{fig:example_automaton});
\item an attempted action blocked by the intelligent tutor (try X event in Figure \ref{fig:example_protocol}), because this action is wrong and the tutoring strategy has been configured to prevent students from performing this action; or
\item an error detected by the intelligent tutor at the time of validating an incorrect action that has not been blocked (fail X event in Figure \ref{fig:example_automaton}). A wrong action may involve more than one error, each of which will be considered as an event.
\end{itemize}

Student logs can also contain irrelevant actions. This type of student actions does not have any influence on the development of the practical assignment. Depending on the pedagogical value of these actions, they will be considered for creating the collective student model (and treated as right actions) or will be discarded.

Accordingly, states represent the different situations derived from the events generated by students. 

Each state $s$ contains the number of students whose logged sequences of events have passed through that state, which is denoted by $\gamma(s)$. The support of a state $s$ is defined as the rate of $\gamma(s)$ with respect to the total number of students in the same cluster. Likewise, each transition $t$ also contains its student frequency, denoted by $\phi(t)$, which is the number of logged sequences of events have passed through that transition. From this frequency, the confidence of the transition $t$ is defined as the rate of $\phi(t)$ with respect to $\gamma(s1)$ where $s1$ is the source state in transition $t$. Figure \ref{fig:example_automaton} denotes support and confidence as percentages in the states and the transitions, respectively.

\subsubsection{Correct Flow Zone} 
This area includes the states that constitute the valid sequences of actions for an exercise, which logically end up with a satisfactory final state. These states are depicted by white circles in Figure \ref{fig:example_automaton}.

\subsubsection{Irrelevant Errors Zone} 
This area groups states derived from error events that do not influence the final result. These error events are associated with attempted actions blocked by the tutor. States in this zone are depicted by yellow circles in Figure \ref{fig:example_automaton}.

As a student may attempt the same wrong action more than once consecutively, vector transitions are employed for outgoing transitions from states of this zone to represent event frequencies. Therefore, the first position of the vector contains the number of students that exit the state without ever repeating the wrong action; the second position contains the number of students that exit the state after repeating the wrong action once and so forth. 

For example, Figure \ref{fig:example_automaton} illustrates the vector transition from yellow state 2 to white state 2 for students that reached state 2 by attempted action 3, where 5\% exited that state without repeating the attempted action ``try 3'' and 15\% exited that state after repeating the attempted action ``try 3'' once.

\subsubsection{Relevant Errors Zone}
This area comprises states derived from error events that actually influence the final result, i.e., if an event of this type occurs the final result will be wrong, unless a repair action is taken. If the error is not repaired, it will be propagated to the subsequent states, which will also be considered erroneous, no matter what the student does afterwards (unless it is a repair action). The states derived directly from relevant errors are depicted by red circles, and the subsequent states that are the result of applying a right action to a state in this zone are denoted by orange circles. 

\subsection{Example Automaton} \label{sec:example}
\begin{figure}[!htb]
 \centering
 \includegraphics[width=\columnwidth]{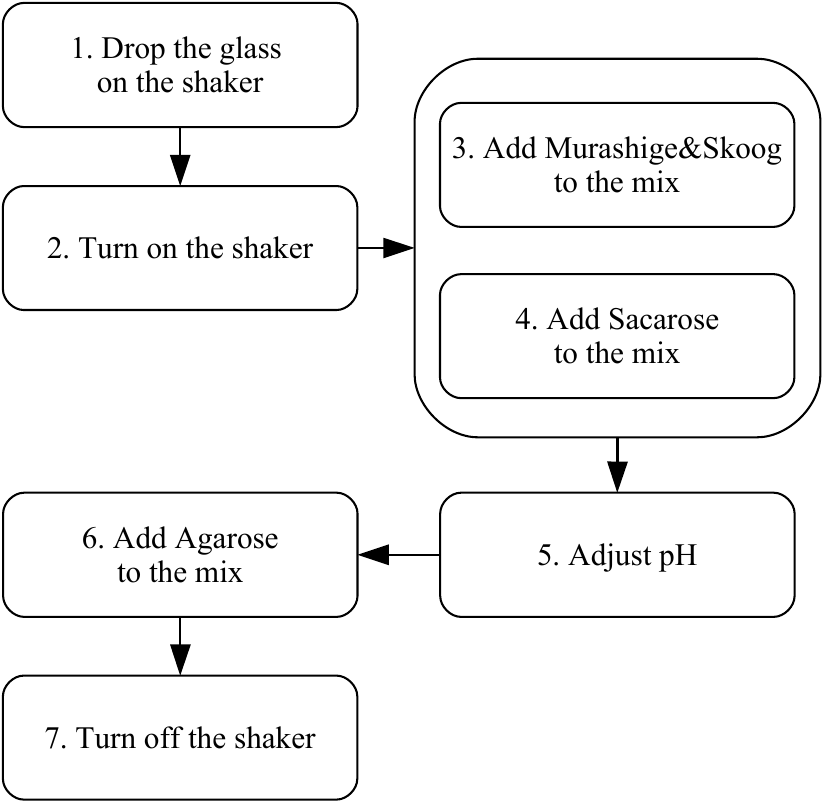}
 \caption{Example of biotechnology lab process}
 \label{fig:example_protocol}
\end{figure}

As an illustrative example of this automaton, we take an excerpt from the biotechnology virtual laboratory protocol \cite{Rico2012}, which is shown in Figure \ref{fig:example_protocol}.

Figure \ref{fig:example_protocol} shows several actions, all of which, except for two (actions 3 and 4), must be executed in a specific order. These two actions can be performed in any order, but both must always be executed before action 5 and after action 2.

Figure \ref{fig:example_automaton} shows an example of the extended automaton described above. This automaton has been created from a subset of the logs of students who have performed the process shown in Figure \ref{fig:example_protocol} within the virtual laboratory for teaching biotechnology. Hence, it reflects all the actions and errors that students have performed in this part of the practical assignment (we have not included the frequencies of all events for the sake of clarity).

Figure \ref{fig:example_automaton} shows how some events associated with right actions recorded in student logs produce correct states. For example, event ``do 1'' (drop the glass on the shaker) leads to state 1, and event ``do 3'' leads to state 3.

In contrast, the event that leads to yellow state 2 does not represent a right action, but the error event of trying to do action 3 instead of action 2. Likewise, yellow state AP (already performed) is the result of another irrelevant error, i.e., trying to do action 4 when it has already been performed previously, because some students performed \(do~4 \to do~3\) from white state 2.

The ``do AC'' event, which is not illustrated in Figure \ref{fig:example_protocol}, represents the action of Adding Casein (a protein) to the mix. This action is considered incompatible with the other actions, but, for pedagogical reasons, the laboratory makes more chemical products available to students than required for the mix. Hence, as soon as the next chemical product is added to the mix (action 3 is performed), this event is validated as incorrect producing a ``fail AC'' event that leads to a red AC state. As Figure \ref{fig:example_automaton} shows, some students completed the remainder of the practical assignment correctly after performing ``do AC'' (path of orange states to orange state 7)

As regards relevant errors, red state 5 is caused by the fail event of not doing the right action 5 next, because some students performed action 6 or action 7 too early. For example, the automaton represents the fact that some students mistakenly performed action 6 instead of action 5 by the path: \(do~6 \to fail~5\). As explained above, a wrong action can cause more than one error event at the same time in a student log. For example, if some students mistakenly perform action 7 instead of action 5, it will be represented by the path: \(do~7 \to fail~5 \to fail~6\), because students have skipped two consecutive right actions.

The frequencies in Figure \ref{fig:example_automaton} show that many students forgot to turn on the shaker before adding substances to the mix (try 3 event labelled with 40\%), or that most students forgot to adjust the mix pH (red state 5 labelled with 70\%). In view of the error rates in these two cases, it might be a good idea to set hints at state 1 and/or the two states 4 if the instructor wants to stop most students from making these mistakes.

As explained previously, a different automaton is built for each cluster. Therefore, the frequencies of this example may not be the same as for the automaton of another cluster. Namely, red state 5 may not have such a high frequency or may not even exist. In this way, the tutoring feedback for students of different clusters may differ.

\subsection{Formal Definition}
Based on the definition of deterministic finite state automata, we define the extended automaton as follows:

\begin{equation}\label{eq:automaton_def}
\begin{aligned}
EA=(&S,S_{C},S_{IE},S_{RE},S_{CE},Z_C,Z_{IE},Z_{RE},\\
&s_0,T_N,T_V,\delta,\gamma,\phi_N,\phi_V)\\
\end{aligned}
\end{equation}
\begin{align*}
\begin{aligned}
Z_C&=S_C\\
Z_{IE}&=S_{IE}\\
Z_{RE}&=S_{RE} \cup S_{CE}\\
S&=S_{C} \cup S_{IE} \cup S_{RE} \cup S_{CE}\\
S_{C} &\cap S_{IE} \cap S_{RE} \cap S_{CE} = \emptyset
\end{aligned}
\end{align*}

In Equation \ref{eq:automaton_def} \(S\) represents the set of all states, and \(Z_C\), \(Z_{IE}\) and \(Z_{RE}\) respectively include the states of the three zones defined above. \(s_0\) is the initial state. \(T_N\) and \(T_V\) are sets of transitions between the states such that \(T_N\) includes normal transitions and \(T_V\) comprises vector transitions. \(\delta\) is the transition function from one state to another, $\gamma$ is a function that associates a state with its frequencies, and $\phi_N$ and $\phi_V$ are functions that associate a normal transition or a vector transition respectively with their frequencies.

\(Z_C\) defines the correct flow zone, which contains correct states (\(S_C\) - white states in Figure \ref{fig:example_automaton}). \(Z_{IE}\) defines the irrelevant errors zone, which contains states of the same type (\(S_{IE}\) - yellow states in Figure \ref{fig:example_automaton}). \(Z_{RE}\) is the relevant errors zone, which, in turn, contains two subsets of states \(S_{RE}\) (red states in Figure \ref{fig:example_automaton}) and \(S_{CE}\) (orange states in Figure \ref{fig:example_automaton}). \(S_{RE}\) is the set of states that are derived directly from relevant errors, whereas \(S_{CE}\) is the set of states that are the result of applying a right action to a state in this zone \(Z_{RE}\); repair actions that do not correct all student mistakes belong to the \(S_{CE}\) group.

The function \(\delta\) is defined in Equation \ref{eq:delta}, where \(s_i\) is the initial state and \(s_f\) is the final state of the given transition.

\begin{equation}
\begin{aligned}\label{eq:delta}
\delta&:T \rightarrow S\\
\delta(t) &= (s_i,s_f)
\end{aligned}
\end{equation}

Function \(\gamma\) is defined for each state in \(S\) by providing the frequency of the state, thereby extending the classic definition.
\begin{equation}
\begin{aligned}\label{eq:gamma}
&\gamma :S \rightarrow \mathbb{N} \\
&\gamma(s) = f
\end{aligned}
\end{equation}

Like the \(\gamma\) function, we define two functions \(\phi_N\) and \(\phi_V\), which provide the frequencies:
\begin{align}
&\begin{aligned}\label{eq:phi_n}
&\phi_N:T_N \rightarrow \mathbb{N} \\
&\phi_N(t) = f
\end{aligned}\\
&\begin{aligned}\label{eq:phi_v}
&\phi_V:T_V \rightarrow \mathbb{N}^n \\
&\phi_V(t) = f^n.
\end{aligned}
\end{align}

\subsection{Build Automaton Process }\label{sec:buildingprocess}
The build automaton process is detailed in Algorithm \ref{alg:automatonbuild}. The inputs of this algorithm are the student logs and a cluster, to which students belong, which it uses to create the automaton for this cluster. 

This algorithm traverses each student log in chronological order and adds each student event in the log to the automaton whose input is the current state $s_i$.

The process of adding an event firstly checks whether there is already a state in the automaton that is the result of this event. If no such state exists, a new state resulting from this event is created, and the current state becomes the new state; otherwise, the frequency of the existing state is increased, and then the process looks for a transition from the current state to the existing state that matches the input event; if such a transition exists, then its frequency is incremented, else a new transition from the current state to the existing state is created. When creating a new transition or increasing its frequency, it is necessary to make a distinction between whether or not it is a vector transition. 

To decide whether a state is a result of an event, or an existing transition matches an event, states and transitions should store some information on their source event. For the sake of clarity, however, we will skip these implementation details in the algorithm specification.

\begin{algorithm}[!b]
  \caption{Build Automaton}
  \label{alg:automatonbuild}
\textbf{Inputs}
\begin{algorithmic}
	\State Input $student~logs$
    \State Input $cluster$
\end{algorithmic}
\textbf{Outputs}
\begin{algorithmic}
	\State Output $EA$ (see Equation \ref{eq:automaton_def})
\end{algorithmic}
  \smallskip
\begin{algorithmic}
\ForEach {\(student \in cluster\)}
    \State \(s_i=s_0\)
  	\State Read \textit{log} of \textit{student} from \textit{student logs}
    \ForEach {\(event \in log\)}
      \State \Call{AddEventLog}{$event,s_i$}
	\EndFor
\EndFor\\
\Procedure{AddEventLog}{$event,s_i$}
	\If {\(\exists s_f \in S | s_f \text{ is result of }event}\)	
      	\State \(\gamma(s_f)=\gamma(s_f)+1\) (see Equation \ref{eq:gamma})
        \If {\parbox[c]{.5\linewidth}{\strut\(\exists t \in model | \delta(t) = (s_i,s_f)  
        \wedge \ t \text{ matches }event\)\strut}}
        	\If {\(t \in T_V\)}
            	\State \(count=\parbox{15em}{\strut number of consecutive events \\equal to \textit{event} in \textit{log}\strut}\)
               \State $\phi_V(t)_{[count]}=\phi_V(t)_{[count]}+1$
               \State (see Equation \ref{eq:phi_v})
     \algstore{alg1}
\end{algorithmic}
\end{algorithm}

\begin{algorithm}[!t]
  \begin{algorithmic}
      \algrestore{alg1}  
                \State \parbox{17em}{\strut Remove consecutive events equal to \textit{event} from \textit{log}\strut}
            \Else            	
                \State \(\phi_N(t)=\phi_N(t)+1\) (see Equation \ref{eq:phi_n})
            \EndIf
        \Else
        	\State \Call{AddTransition}{$s_i,s_f$}
        \EndIf
        \State \(s_i=s_f\)       
      \Else 
    	\State Add new state \(s \text{ to } S\) ($S$ is element of $EA$).
        \State \Call{AddTransition}{$s_i,s$}
        \State \(s_i=s\)
      \EndIf
\EndProcedure\\

\Procedure{AddTransition}{$s_i,s_f$}	
	\If {$s_i \in S_{IR}$} (see Equation \ref{eq:delta})
		\State Add a new \(t_V\) to \(T_V\) from \(s_i\) to \(s_f\) using $\delta(t)$
        \State ($T_V$ is element of $EA$)
    \Else
		\State Add a new \(t_N\) to \(T_N\) from \(s_i\) to \(s_f\) using $\delta(t)$
        \State ($T_N$ is element of $EA$)
    \EndIf
\EndProcedure
\end{algorithmic}
\end{algorithm}

\subsection{Application of the Collective Student Model in the Tutoring Process}\label{sec:application_model}
As mentioned above, the proposed ITS relies on the collective student model to stop students from making mistakes or from floundering with the practical assignment. For this purpose, the ITS can use the collective student model at training time to detect a situation in which there is a significant probability of the student making a mistake or getting stuck. If it is pedagogically inconvenient for the student to be allowed to make the mistake, the ITS will have to be configured to warn the student by means of a hint.

The simplest case in which the model can indicate that the student is highly likely to make a mistake is when the confidence in a transition from the student's current state to an error state is high. For example, ``try 3'' from state 1 in Figure \ref{fig:example_automaton}. The instructor responsible for the practical assignment will specify when a confidence or support is high.

However, the above is just one particular case of a more general situation where an error state is highly likely to be reached after several consecutive transitions from the current student state. This situation is referred to as an indirect transition. It is important to identify such situations when the support for a relevant error state is high. For example, if the current state of the student in Figure \ref{fig:example_automaton} is white 2, the error state AC could be reached through the sequence of transitions \(do~AC \to do~3 \to fail~AC\). Besides, this is a highly likely sequence, because 60\% of students who reach white state 2 (all of the students eventually reach white state 2) get the red state AC.

When working out the probability of reaching an error state through an indirect transition, we may take into account more than one alternative path in the automaton. For example, let us consider red state 5 as the error state in Figure \ref{fig:example_automaton}, then it is clear that there are three paths to red state 5 from white state 3: \(fail~AC \to do~4 \to do~6 \to fail~5 \), \(do~4 \to do~6 \to fail~5 \) and \(do~4 \to do~7 \to fail~5 \). Besides, as can be deduced from the automaton, 70\% of students at white state 3 will eventually reach red state 5, because all the students reach white state 3 at some point (nobody forgot to do action 3).

In order to speed up the real-time performance of the ITS, indirect transitions may be computed offline and included in the automaton as direct transitions.

A student is said to be floundering when he or she repeats the same wrong action many times in the practical assignment. To prevent students from getting stuck, and probably getting frustrated, we need to find out whether there is a high probability of the student proceeding from the current state to an irrelevant error state where he or she may repeat a wrong action several times. To do this, the irrelevant error state must have an outgoing vector transition with the right content, as applies in yellow state 2 with vector transition ``do 2'' in which 70\% of students will exit that state after repeating the wrong action ``try 3'' three times.

\subsubsection{Update model process}
The collective student model is updated with each new event executed by each student at training time. The update process searches the state that is the result of the last event executed by a student. From this state, it creates a new or increments the frequency of an existing transition in a similar fashion to the build automaton process.

The process for updating the model is as follows:
\begin{enumerate}
\item Search for the current state $s_i$ that represents the last event executed. If it is the beginning of the practical assignment, the current state will be state 0.

\item Execute the \textit{AddEventLog} procedure from Algorithm \ref{alg:automatonbuild}, where $event$ is the next event to be processed and $s_i$ is the current state.
\end{enumerate}

\subsubsection{Student reclassification}\label{section:reclassification}
When a student has just started a practical assignment, he or she is assigned to a default cluster. The default cluster is the one that has the average values of the attributes used to generate the clusters.

As the student progresses in the practical assignment, however, the initial classification may no longer hold. It is the instructor that initially configures the checkpoints determining when to check whether a student no longer belongs to a cluster. These checkpoints may either be particular steps in the practical assignment protocol or be defined by percentages of the average total time taken by the students in the current cluster. Therefore, when the student reaches any of these checkpoints, he or she can be reclassified according to his or her log so far. To do this, we apply the following process:

\begin{enumerate}
\item Get the current cluster to which the student belongs.
\item Select the cluster that best matches the student log.
\item If the best match is the same cluster, do nothing.
\item Else
  \begin{enumerate}
  \item Undo the application of the student log in the automaton of the former cluster by doing the opposite to what Algorithm \ref{alg:automatonbuild} does (decreasing frequencies and/or removing states).
  \item Apply the student log to the automaton of the new cluster according to Algorithm \ref{alg:automatonbuild}.
  \end{enumerate}
\end{enumerate}

In order to select the cluster that best matches the student log at a checkpoint, we can use the same method as for the initial clustering, albeit considering just the excerpts of the logs of students that have reached the same checkpoint as the log of the student under supervision. This selection consist of searching the cluster where its centroid is close to the value of the attributes calculated from the student log, technique similar to that used in incremental clustering \cite{Rokach2005}.

\section{Collective Student Model Validation}\label{sec:model_val}
This validation has two main objectives. The first goal is to verify that the error in the predictions computed by the proposed model is acceptable for automatic tutoring. Roughly, the error associated with the prediction of a student action is equal to 1 minus the probability of the student performing that action according to the model. Hence, if the model predicts that there is a probability of 1 that the student will perform an action and the student eventually performs that action, the error will be 0. Otherwise, if the student does not perform that action, the error will be 1. The second goal is to check whether clustering methods can divide students into groups that require different tutoring feedback. To do this, we will compare the clusters by checking how many students in each cluster made a relevant error in the practical assignment.

In order to validate the proposed model, we will employ the student logs taken from a virtual laboratory for teaching biotechnology \cite{Rico2012}. This laboratory is an educational 3D virtual environment for procedural training used by students to carry out a practical assignment composed of around 120 actions, such as add a chemical to a mix or turn on a machine. This virtual laboratory was implemented on the OpenSimulator platform.

Student logs were collected by a component inside this virtual laboratory, called automatic tutor. This component validates student actions and registers events like the ones outlined in Section \ref{sec:example}.

\subsection{Method}\label{sec:valmethods}
For the last three years (2013, 2014 and 2015), 85 students taking the Biochemistry and Biotechnology course taught as part of the BSc in Forestry Engineering at the UPM used the biotechnology virtual laboratory \cite{Rico2012} to complete a practical assignment. The automatic tutor of this virtual lab collected and classified the events that these students generated during this practical assignment. These events may fall into any of the following categories: 
\begin{itemize}
\item Correct Events
  \begin{itemize}
  \item Corrective Events
  \item Non-Corrective Events
  \end{itemize}
\item Error Events
  \begin{itemize}
  \item Dependency Errors
  \item Incompatibility Errors
  \item World Errors
  \item Other Errors.
  \end{itemize}
\end{itemize} 

Correct events are right actions. However, when a right action repairs an error made previously, the associated event is considered to be a corrective event. Dependency or incompatibility errors depend on the configuration of the virtual laboratory (detailed in \cite{Rico2012}) set up by the instructor. They are related to the right order in which to perform the actions in the practical assignment. World errors refer to failures in the handling of 3D objects, for example, if a student tries to drop an object where it should not be dropped. Finally, the other error events category represent errors that are not pedagogically relevant, for example, if the student tries to repeat an action that has already been performed.

\subsubsection{Model mean error calculation}
To calculate the model mean error, we proceeded as follows:
\begin{enumerate}
\item \label{itm_validateproc:firsstep} Select 90\% of students at random.
\item Apply the process defined in Section \ref{sec:buildingprocess} to build a model using the logs from the selected 90\% of students.
\item \label{itm_validateproc:laststep} Validate how well the model aligned with the remaining 10\% of student logs.
\item Repeat step \ref{itm_validateproc:firsstep} to \ref{itm_validateproc:laststep}   
 \textit{n} times to guarantee the representativeness of the randomly selected students.
\end{enumerate}

The 10\% of students used in step \ref{itm_validateproc:laststep} constitute the \textit{test set} for the model. This test set is used to estimate how accurately the model will perform in the practical assignment, like many validation algorithms in data mining or statistics.

The number $n$ of times that steps 1 to 3 are repeated was 150 after checking that with $n=75, 100 \text{ and } 150,$ results were equivalent.

The process for validating the alignment (step \ref{itm_validateproc:laststep}) of the test set with the model is defined in Algorithm \ref{alg:validationproc}. The inputs of this algorithm are the model, the confidence and the support that determine the submodel to be validated, and the test set of students. For example, if \(confidence=support=0\), then the entire model will be validated.

\begin{algorithm}[!t]
  \caption{Validation process}
  \label{alg:validationproc}
  \textbf{Inputs}
\begin{algorithmic}
	\State Input $model$
	\State Input $support$
    \State Input $confidence$
    \State Input $val\_example\_set$
\end{algorithmic}
\textbf{Outputs}
\begin{algorithmic}
	\State Output $\bar{E}$
\end{algorithmic}
  \smallskip
\begin{algorithmic}
\ForEach {\(student\_log \in val\_example\_set\)}
	\State Get \(cluster \in model | student\_log \text{ best fits } cluster\)
    \State Get the \(automaton \in cluster\)
    \State Get \(s_0 \in automaton\)
    \State \(previous\_state = s_0\)

    \ForEach {\(event \in student\_log\)}
        \If {\parbox[c]{.66\linewidth}{\strut\(\exists state \in automaton | state \text{ is the result of } event\)\strut}}
    		\State \parbox[c]{.6\linewidth}{\strut\(exist(t) \equiv  \exists t \in automaton | \delta(t) = (previous\_state, state)\)\strut} 
             \If {\( exist(t)\)}
               \If {\parbox[c]{.4\linewidth}{\strut\(state_{sup} \geq support \wedge t_{conf} \geq confidence\)\strut}}
                    \State \parbox[c]{.5\linewidth}{\strut Calculate \(E_{event}\) according to equations \ref{eq:E}, \ref{eq:EV} or \ref{eq:EL}}
                \Else
                 \State \(E_{event} = 0\)
                \EndIf
              \Else
        			\State \(E_{event} = 1\)
        	  \EndIf  
        \Else
        	\State \(E_{event} = 1\).
        	\State \(state = \text{new temporary state from } event\)
        \EndIf
        \State \(previous\_state = state\)
	\EndFor
    \State Update \(\bar{E} \text{ with } E_{event}\) according to equation \ref{eq:EM}
\EndFor
\end{algorithmic}
\end{algorithm}

Temporary states in Algorithm \ref{alg:validationproc} are states that are not in the model but have to be created temporarily for validation purposes. For instance, if a student in the validation test set generates an event, and the model does not contain any state that is the result of that event, the algorithm creates a new state which it links to the previous state (this new state is not added to the model). This operation is performed until the next event in the log matches a state in the model.

The alignment error of an event log with the model is calculated using a similar equation to Replay Fitness \cite{Buijs2014}, which is employed to quantify the extent to which the model can reproduce the traces recorded in the log in process discovery. Equation \ref{eq:E} defines this calculation for transitions with normal frequencies (\(\in T_N\)), where \textit{t} is the transition that matches the event performed by the student and \textit{s} is the initial state of \textit{t} (functions $\phi_N$ and $\gamma$ are defined in equations \ref{eq:phi_n} and \ref{eq:gamma} respectively).

\begin{equation}\label{eq:E}
E_{event}=1-\frac{\phi_N(t)}{\gamma(s)}
\end{equation}

There is a slightly different equation for calculating this error in the case of a vector transition (\(\in T_V\)). If the student exits the loop the first time round, a similar equation to the above is used to calculate the error, where the transition frequency is replaced with the frequency stored in the first element of the vector (function $\phi_V$ is defined in equation \ref{eq:phi_v}).

\begin{equation}\label{eq:EV}
E_{event}=1-\frac{\phi_V(t)[1]}{\gamma(s)}
\end{equation}

If the student does not exit the loop first time round, the transition frequency is replaced by the sum of all elements of the frequency vector, except from the first, i.e.,

\begin{equation}\label{eq:EL}
E_{event}=1-\frac{\sum^{\phi_V(t)_{count}}_{i=2} \phi_V(t)[i]}{\gamma(s)}.
\end{equation}

The reason for making a distinction between these two cases when calculating the vector transition error is that it makes more sense, when providing tutorial hints, to work with different predictions depending on whether a student makes a loop error once or more than once. For example, if a student is more likely to make a mistake only once rather than more than once, it will be less necessary to display a hint to avoid that mistake.

To calculate the mean event error, each event error $E_i$ calculated by the above equations is multiplied by the number of students in the test set that generated the same event $n_i$, and the sum of these products is divided by the sum of the number of students that generated each event.

\begin{equation}\label{eq:EM}
\bar{E}=\frac{\sum^{n~dif~events}_{i=1} E_{i}n_{i}}{\sum^{n~dif~events}_{i=1} n_{i}}
\end{equation}

\subsubsection{Clustering methods}\label{sec:clumet}
As selecting the clustering methods \cite{Rokach2005,Han2012Ch10,Han2012Ch11,Tan2013Ch8}, we only considered methods that do not need initially the $k$ number of clusters. Thus, we selected the following methods:

\begin{enumerate}
\item XMeans
\item Expectation Maximization (EM)
\item Microsoft Sequence Clustering
\item No clustering.
\end{enumerate}

The first two are partitioning methods, because they construct the groups based on the distances that exist between the objects. These methods use an iterative relocation technique, which moves each object into a group by analyzing how close is to other members of the group and how far is from objects in other groups. This distance measure is calculated by means of the Euclidean distance between two vectors.

The functions that calculate the vector associated with an object are as follows:

\begin{enumerate}
\item Clustering by errors
\item Clustering by errors and time
\item Clustering by events in each zone of the automaton.
\end{enumerate}

The first function returns an error coefficient, which is calculated using the weighted sum of errors made by a student with a different weight depending on the pedagogical importance of the error. The second function returns an error coefficient-time pair, where the error coefficient is calculated as explained before, and time stands for the time that it took the student to complete the entire practical assignment. If the student did not complete the practical assignment, he or she is penalized with 24 hours. The clustering by events in zone function returns a triple with the number of events of each type existing in a student log.

The Microsoft Sequence Clustering algorithm \cite{Microsoft} uses Markov chain analysis to identify ordered sequences. Then clusters are generated from these results using an Expectation Maximization algorithm applied to the ordered sequences. We ruled out the Weka Sequential Information Bottleneck algorithm (sIB) \cite{Slonim:2002} because it requires the number of clusters as input.

For the first two methods (XMeans and EM), we used their implementation in the Weka framework \cite{Hall2009}. 

\subsection{Results}
To get an idea of the data size, we built an automaton without clustering the logs of the 85 students (14943 log events). The result was an automaton with 532 states and 2778 transitions (1454 normal and 1324 vector). The number of states is detailed in Table \ref{table:statesbyzone} for each zone of this automaton.

\begin{table}[h]
\caption{States by zone in a non-clustered model for validation}
\label{table:statesbyzone}
\centering
\begin{tabular}{ c | c }
\hline
\textbf{Zone} & \textbf{Number of States}\\
\hline
Correct Flow & 115\\ 
Irrelevant Errors & 228\\
Relevant Errors & 189\\
\hline
\end{tabular}
\end{table}

\subsubsection{Model mean error}
The support and confidence values 0, 0.1, 0.25, 0.5, 0.75, 0.9 were used as inputs for the validation process, since this set of values was considered representative and complete enough to illustrate the error evolution.

Table \ref{table:errorbyconfsupp} shows the model mean error for the best method and function combination and all the confidence and support combinations. Model mean error was calculated using the process explained in Section \ref{sec:valmethods}.

\begin{figure}[!h]
\centering
	\includegraphics[width=\columnwidth]{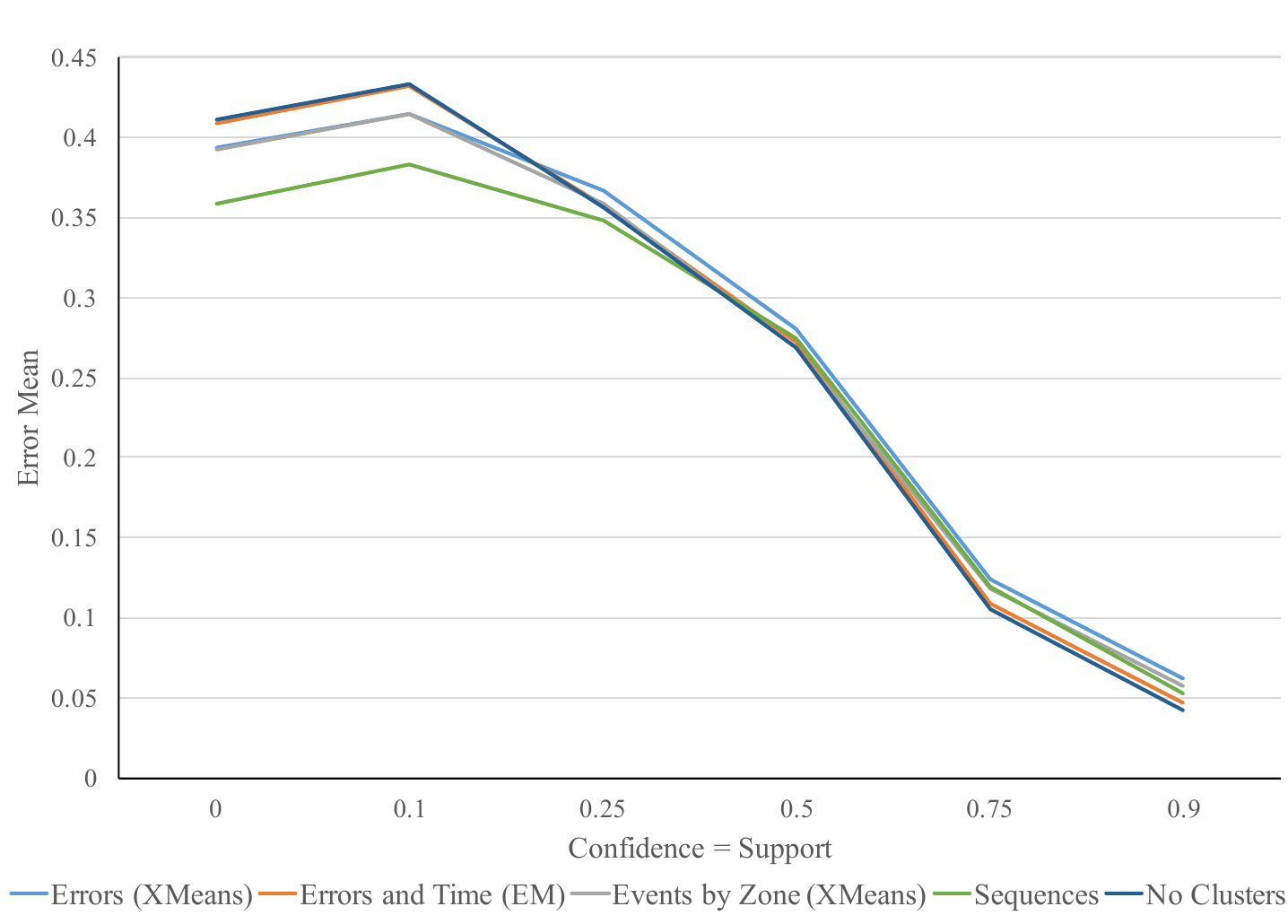}
	\caption{Mean error when \(confidence=support\)}
 	\label{fig:errormeanconfeqsup}
\end{figure}

Figure \ref{fig:errormeanconfeqsup} highlights that the differences between functions are very small (about 0.01) around 0.5. Also, at some point between 0.5 and 0.75, the no clustering line drops below all the others. 

\begin{figure}[!h]
\centering
	\includegraphics[width=\columnwidth]{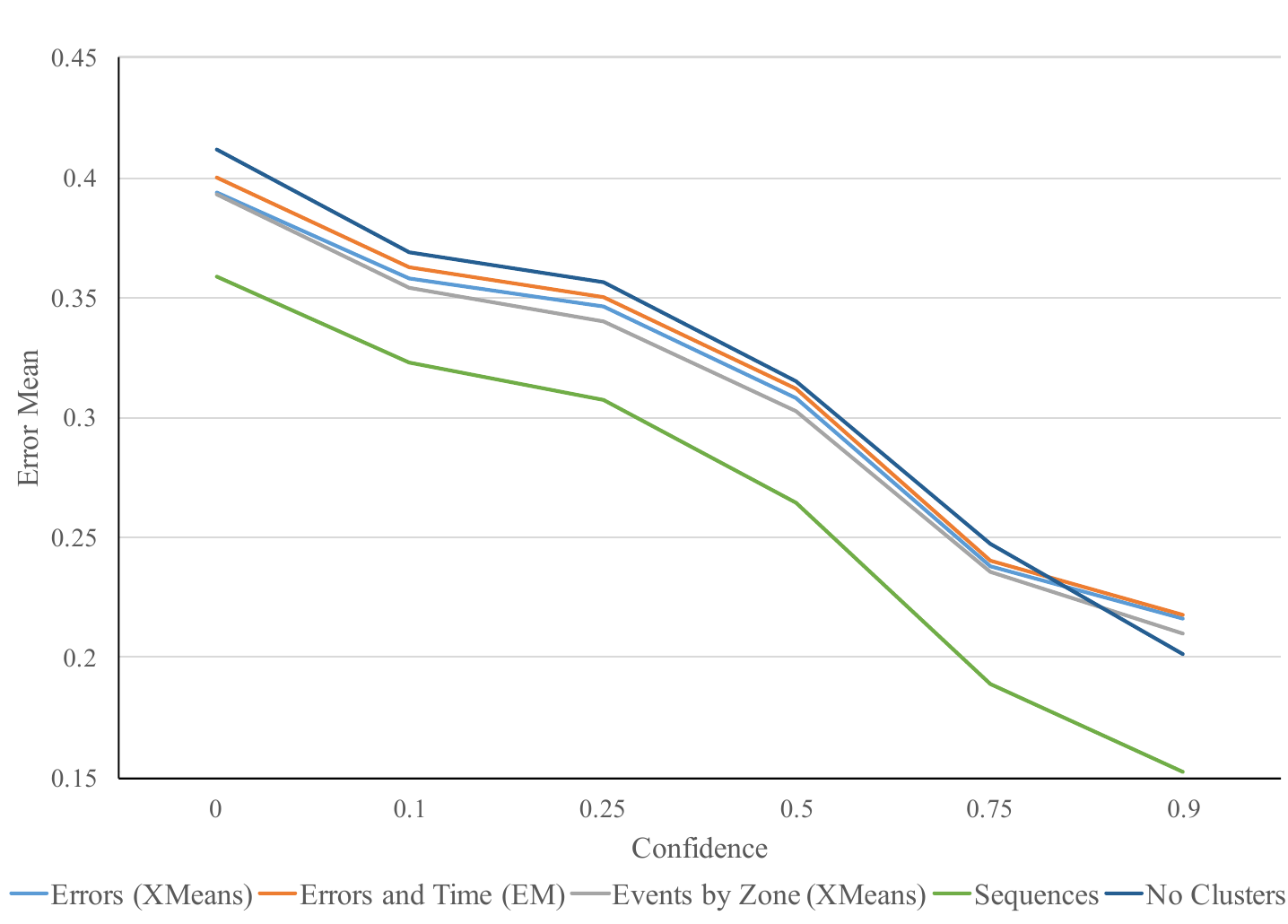}
	\caption{Mean error when \(support=0\)}
 	\label{fig:errormeansup0}
\end{figure}

As explained in Section \ref{sec:application_model}, we are interested for tutoring purposes in transitions that are highly likely to reach an error state. Conceptually, this is equivalent to working with a submodel with a high confidence (for example, \(confidence=0.5\)) and \(support=0\). Figure \ref{fig:errormeansup0} shows model mean error with a confidence of between 0 and 0.9 and \(support=0\).

\begin{table*}[!ht]
\caption{Mean error by method and clustering function.}
\label{table:errorbyconfsupp}
\centering
\begin{tabular}{l@{ }l@{}c@{ }l@{ } | llllll}
\hline
\textbf{Function} & \textbf{Method} & \begin{tabular}{@{}c@{}}\textbf{Num.} \\ \textbf{of} \\ \textbf{Clust.}\end{tabular} & \begin{tabular}{@{}l@{}} \ \\ \ \ {\scriptsize \textbf{Sup.}}\\{\scriptsize \textbf{Conf.}}
\end{tabular} & \textbf{0}     & \textbf{0.1}   & \textbf{0.25}  & \textbf{0.5}   & \textbf{0.75}  & \textbf{0.9}   \\
\hline
\multirow{6}{*}{Errors}	& \multirow{6}{*}{XMeans}	& \multirow{6}{*}{2} & 0        & 0.394 & 0.358 & 0.346 & 0.308 & 0.238 & 0.217 \\
                                 &                                                &                    & 0.1      & 0.483 & 0.415 & 0.388 & 0.312 & 0.162 & 0.131 \\
                                 &                                                &                    & 0.25     & 0.463 & 0.389 & 0.366 & 0.285 & 0.126 & 0.069 \\
                                 &                                                &                    & 0.5      & 0.416 & 0.330 & 0.323 & 0.281 & 0.124 & 0.062 \\
                                 &                                                &                    & 0.75     & 0.361 & 0.213 & 0.208 & 0.198 & 0.124 & 0.062 \\
                                 &                                                &                    & 0.9      & 0.435 & 0.117 & 0.109 & 0.097 & 0.088 & 0.062 \\
\hline
\multirow{6}{*}{\begin{tabular}{@{}l@{}}Errors \\ and \\ Time \end{tabular}} & \multirow{6}{*}{EM} & \multirow{6}{*}{2} & 0        & 0.400 & 0.363 & 0.350 & 0.312 & 0.240 & 0.218 \\
                                 &                                                &                    & 0.1      & 0.493 & 0.423 & 0.394 & 0.312 & 0.159 & 0.145 \\
                                 &                                                &                    & 0.25     & 0.473 & 0.397 & 0.370 & 0.285 & 0.123 & 0.080 \\
                                 &                                                &                    & 0.5      & 0.422 & 0.331 & 0.323 & 0.277 & 0.120 & 0.070 \\
                                 &                                                &                    & 0.75     & 0.372 & 0.212 & 0.207 & 0.194 & 0.120 & 0.070 \\
                                 &                                                &                    & 0.9      & 0.452 & 0.117 & 0.106 & 0.092 & 0.083 & 0.070 \\
\hline
\multirow{6}{*}{\begin{tabular}{@{}l@{}}Events \\ by \\ Zone \end{tabular}}  & \multirow{6}{*}{XMeans} & \multirow{6}{*}{2} & 0        & 0.393 & 0.354 & 0.340 & 0.303 & 0.236 & 0.210 \\
                                 &                                                &                    & 0.1      & 0.485 & 0.414 & 0.383 & 0.302 & 0.154 & 0.140 \\
                                 &                                                &                    & 0.25     & 0.465 & 0.389 & 0.358 & \cellcolor[RGB]{155,194,230}0.274 & 0.119 & 0.059 \\
                                 &                                                &                    & 0.5      & 0.414 & 0.321 & 0.309 & \cellcolor[RGB]{174,170,170}0.269 & 0.119 & 0.057 \\
                                 &                                                &                    & 0.75     & 0.367 & 0.207 & 0.202 & 0.190 & 0.119 & 0.057 \\
                                 &                                                &                    & 0.9      & 0.444 & 0.112 & 0.104 & 0.092 & 0.085 & 0.057 \\
\hline
\multirow{6}{*}{Sequences}       & \multirow{6}{*}{\begin{tabular}{@{}l@{}}Microsoft \\ Sequence \\ Clustering \end{tabular}} & \multirow{6}{*}{6} & 0        & \cellcolor[RGB]{221,235,247}0.358 & \cellcolor[RGB]{221,235,247}0.323 & \cellcolor[RGB]{221,235,247}0.307 & \cellcolor[RGB]{221,235,247}0.264 & \cellcolor[RGB]{221,235,247}0.189 & \cellcolor[RGB]{221,235,247}0.153 \\
                                 &                                                &                    & 0.1      & \cellcolor[RGB]{224,220,220}0.445 & \cellcolor[RGB]{224,220,220}0.383 & \cellcolor[RGB]{224,220,220}0.361 & \cellcolor[RGB]{224,220,220}0.290 & \cellcolor[RGB]{224,220,220}0.146 & 0.123 \\
                                 &                                                &                    & 0.25     & \cellcolor[RGB]{155,194,230}0.427 & \cellcolor[RGB]{155,194,230}0.365 & \cellcolor[RGB]{155,194,230}0.349 & 0.280 & 0.136 & 0.105 \\
                                 &                                                &                    & 0.5      & \cellcolor[RGB]{174,170,170}0.377 & \cellcolor[RGB]{174,170,170}0.306 & 0.300 & 0.274 & 0.120 & 0.053 \\
                                 &                                                &                    & 0.75     & \cellcolor[RGB]{132,151,176}0.324 & 0.201 & 0.197 & 0.192 & 0.120 & 0.053 \\
                                 &                                                &                    & 0.9      & \cellcolor[RGB]{47,117,181}0.341 & \cellcolor[RGB]{47,117,181}0.100 & 0.093 & 0.084 & \cellcolor[RGB]{47,117,181}0.066 & 0.053 \\
\hline
\multicolumn{3}{l}{\multirow{6}{*}{No Clustering}}                                                       & 0        & 0.412 & 0.369 & 0.356 & 0.315 & 0.247 & 0.202 \\
                                 &                                                &                    & 0.1      & 0.522 & 0.433 & 0.400 & 0.312 & 0.146 & \cellcolor[RGB]{224,220,220}0.107 \\
                                 &                                                &                    & 0.25     & 0.504 & 0.400 & 0.356 & 0.277 & \cellcolor[RGB]{155,194,230}0.105 & \cellcolor[RGB]{155,194,230}0.042 \\
                                 &                                                &                    & 0.5      & 0.444 & 0.312 & \cellcolor[RGB]{174,170,170}0.298 & 0.269 & \cellcolor[RGB]{174,170,170}0.105 & \cellcolor[RGB]{174,170,170}0.042 \\
                                 &                                                &                    & 0.75     & 0.427 & \cellcolor[RGB]{132,151,176}0.197 & \cellcolor[RGB]{132,151,176}0.186 & \cellcolor[RGB]{132,151,176}0.182 & \cellcolor[RGB]{132,151,176}0.105 & \cellcolor[RGB]{132,151,176}0.042 \\
                                 &                                                &                    & 0.9      & 0.574 & 0.105 & \cellcolor[RGB]{47,117,181}0.085 & \cellcolor[RGB]{47,117,181}0.077 & 0.070 & \cellcolor[RGB]{47,117,181}0.042\\
\hline
\end{tabular}
\par
\vspace{1em}
{\footnotesize Label:
\begin{tabular}{ll}
\cellcolor[RGB]{221,235,247} & Min values with \(sup.=0\)\\
\cellcolor[RGB]{224,220,220} & Min values with \(sup.=0.1\)\\
\cellcolor[RGB]{155,194,230} & Min values with \(sup.=0.25\)\\
\end{tabular}
\begin{tabular}{ll}
\cellcolor[RGB]{174,170,170} & Min values with \(sup.=0.5\)\\
\cellcolor[RGB]{132,151,176} & Min values with \(sup.=0.75\)\\
\cellcolor[RGB]{47,117,181} & Min values with \(sup.=0.9\)\\
\end{tabular}
}
\end{table*}

As Figure \ref{fig:errormeansup0} shows, all clustering functions, and especially the sequences function, have lower error values than the no clustering function. Figure \ref{fig:errormeansup0} also shows that the error decreases rapidly when the confidence for a support of zero increases.

\subsubsection{Errors by cluster}
As explained above, the principal application of the collective student model is to generate anticipated feedback for students. Thus, let us analyze how well these methods group together students according to their performance in the practical assignment.

To do this, we considered the frequencies of the most common pedagogical errors in the clusters output by each clustering method. For this purpose, we took some error states located in the Relevant Errors Zone. Then, for each clustering method, we compared the frequencies of these errors in each cluster.

From this analysis, we selected results output by the events by zone clustering method (Table \ref{tab:eventzone_freqs}) and sequence clustering method (Table \ref{tab:sequence_freqs}). The first column in Tables \ref{tab:eventzone_freqs} and \ref{tab:sequence_freqs} represents the error code; the last column includes the variances of the error frequencies; and the intermediate columns contain the frequencies of the errors in each cluster.

\begin{table}[!b]
\centering
\caption{Frequencies by cluster for events by zone clustering function (XMeans)}
\label{tab:eventzone_freqs}
\begin{tabular}{l|ll|l}
\hline
\textbf{Error}      & \textbf{Cluster 1}   & \textbf{Cluster 2} & \textbf{Variance}\\
\hline
1.20\_1.16 & \cellcolor[RGB]{155,194,230}0.545     & \cellcolor[RGB]{132,151,176}1  & 0.103\\
1.20\_1.14 & 0.364     & \cellcolor[RGB]{132,151,176}0.716  & 0.062\\
2.52\_2.40 & 0.182     & \cellcolor[RGB]{155,194,230}0.649  & 0.109\\
2.52\_2.38 & 0.091     & \cellcolor[RGB]{221,235,247}0.581  & 0.120\\
2.52\_2.37 & 0.091     & \cellcolor[RGB]{221,235,247}0.581  & 0.120\\
2.52\_2.39 & 0.091     & \cellcolor[RGB]{221,235,247}0.568  & 0.114\\
2.52\_2.35 & 0.091     & \cellcolor[RGB]{221,235,247}0.554  & 0.107\\
3.52\_3.34 & 0.182     & \cellcolor[RGB]{221,235,247}0.527  & 0.059\\
2.52\_2.41 & 0.091     & \cellcolor[RGB]{221,235,247}0.527  & 0.095\\
2.52\_2.34 & 0.091     & 0.459 								& 0.068\\
\hline
Average	   & 0.182	   & 0.616								& 0.096\\
\hline
\end{tabular}
\par
\vspace{1em}
{\footnotesize Label:
\begin{tabular}{ll}
  & Frequencies under 50\%\\
\cellcolor[RGB]{221,235,247} & Frequencies between 50\% and 60\%\\
\cellcolor[RGB]{155,194,230} & Frequencies between 60\% and 70\%\\
\cellcolor[RGB]{132,151,176} & Frequencies over 70\%\\
\end{tabular}
}
\end{table}

\begin{table}[!t]
\centering
\caption{Frequencies by cluster for sequences clustering function}
\label{tab:sequence_freqs}
\begin{tabular}{l|llllll|l}
\hline
\textbf{Error} & \textbf{Cl. 1} & \textbf{Cl. 2} & \textbf{Cl. 3} & \textbf{Cl. 4} & \textbf{Cl. 5} & \textbf{Cl. 6} & \textbf{Var.}\\
\hline
1.20\_1.16 & \cellcolor[RGB]{155,194,230}0.688     & \cellcolor[RGB]{132,151,176}0.833     & \cellcolor[RGB]{132,151,176}0.857     & \cellcolor[RGB]{221,235,247}0.6     & \cellcolor[RGB]{221,235,247}0.5     & \cellcolor[RGB]{132,151,176}0.778 & 0.020\\
1.20\_1.14 & \cellcolor[RGB]{132,151,176}0.750     & \cellcolor[RGB]{155,194,230}0.667     & \cellcolor[RGB]{221,235,247}0.5     & \cellcolor[RGB]{155,194,230}0.7     & \cellcolor[RGB]{132,151,176}0.750     & \cellcolor[RGB]{221,235,247}0.556 & 0.011\\
2.52\_2.40 & \cellcolor[RGB]{221,235,247}0.563     & \cellcolor[RGB]{132,151,176}0.750     & \cellcolor[RGB]{221,235,247}0.571     & 0.3     & \cellcolor[RGB]{132,151,176}1     & 0.444 & 0.060\\
2.52\_2.38 & 0.469     & \cellcolor[RGB]{221,235,247}0.583     & \cellcolor[RGB]{221,235,247}0.571     & \cellcolor[RGB]{221,235,247}0.5     & \cellcolor[RGB]{155,194,230}0.625     & 0.444 & 0.005\\
2.52\_2.37 & \cellcolor[RGB]{221,235,247}0.531     & \cellcolor[RGB]{221,235,247}0.583     & 0.429     & \cellcolor[RGB]{221,235,247}0.5     & \cellcolor[RGB]{132,151,176}\cellcolor[RGB]{132,151,176}0.750     & 0.333 &  0.020\\
2.52\_2.39 & \cellcolor[RGB]{221,235,247}0.5     & \cellcolor[RGB]{221,235,247}0.583     & \cellcolor[RGB]{221,235,247}0.5     & 0.3     & \cellcolor[RGB]{132,151,176}0.750     & 0.444 & 0.022\\
2.52\_2.35 & 0.469     & \cellcolor[RGB]{221,235,247}0.583     & \cellcolor[RGB]{221,235,247}0.5     & 0.4     & \cellcolor[RGB]{155,194,230}0.625     & 0.444 & 0.007\\
3.52\_3.34 & \cellcolor[RGB]{221,235,247}0.531     & 0.417     & 0.429     & \cellcolor[RGB]{221,235,247}0.6     & 0.375     & 0.444 & 0.007\\
2.52\_2.41 & 0.375     &\cellcolor[RGB]{221,235,247}0.583     & \cellcolor[RGB]{221,235,247}0.5     & 0.4     & \cellcolor[RGB]{132,151,176}0.750     & 0.444 & 0.020\\
2.52\_2.34 & 0.344     & \cellcolor[RGB]{221,235,247}0.583     & \cellcolor[RGB]{221,235,247}0.5     & 0.3     & \cellcolor[RGB]{155,194,230}0.625     & 0.222 & 0.027\\
\hline
Average & 0.522 & 0.617 & 0.536 & 0.460 & 0.675 & 0.456 & 0.020 \\
\hline
\end{tabular}
\par
\vspace{1em}
{\footnotesize Label:
\begin{tabular}{ll}
  & Frequencies under 50\%\\
\cellcolor[RGB]{221,235,247} & Frequencies between 50\% and 60\%\\
\cellcolor[RGB]{155,194,230} & Frequencies between 60\% and 70\%\\
\cellcolor[RGB]{132,151,176} & Frequencies over 70\%\\
\end{tabular}
}
\end{table}

\hfill \break
Sequence clustering results were selected because this method was the best predictor of student behavior according to Table \ref{table:errorbyconfsupp}. The events by zone clustering method was selected because it proved to be the best at grouping students according to their performance in the practical assignment.

The above two tables show that the variances in Table \ref{tab:eventzone_freqs} are greater than in Table \ref{tab:sequence_freqs}. This difference is clearer in the last row of each table (average frequencies).

\subsection{Discussion}
As expected, Table \ref{table:errorbyconfsupp} and Figure \ref{fig:errormeanconfeqsup} show that errors tend to decrease when confidence and support increase. This is because the most frequent transitions and the most visited states reflect the most common behavior of the students in the practical assignment. Therefore, the predictions based on such transitions and states are more reliable.

We think that the min error (0.358 for the sequences method) yielded by \(support=0\) and \(confidence=0\) is reasonably good for a training environment like this, where students have a lot of freedom of action. Besides, we should take into account that the ITS will typically consider transitions that are likely to reach error states (with a confidence greater than 0.5, for example) for tutoring tasks. Therefore, prediction errors will be lower (less than 0.27 in this experiment) than when $confidence=support=0$. In addition, as explained above, the model is updated with each new student action at training time. This should help to reduce prediction error because the model will tend to fit the students under supervision better.

Another conclusion that we can draw from the results in Table \ref{table:errorbyconfsupp} is that the differences between clustering methods are small. For example, with \(support=0\) and \(confidence=0\) (i.e., the entire model), the min error is 0.358 (\textit{Sequence clustering}) and the max error (\textit{No clustering}) is 0.412. Although these differences are small, the best clustering method is clearly sequence clustering, mostly with \(support=0\), as shown in Figure \ref{fig:errormeansup0}. This supports the approach of using clustering, because, as mentioned above, tutoring tasks will consider transitions that are highly like to reach error states with any support.

Concerning results on errors by cluster, Table \ref{tab:eventzone_freqs} shows that clustering methods, and particularly events by zone clustering, can successfully classify students into groups that will need different tutoring feedback throughout the practical assignment.

Let us take the example of the error in the second row (1.20\_1.14). For this error, cluster 1 in Table \ref{tab:eventzone_freqs} has a low frequency (36.4\%), whereas cluster 2 has a very high frequency (71.6\%). These figures imply that students in cluster 1 should not need any advice on this error because very few former students made that error. In contrast, most of the students in cluster 2 will need a hint to prevent them from making this error, as explained in Section \ref{sec:application_model}.

Taking into account the results of the two studies (model mean error and errors by cluster), the best clustering method for this data set is events by zone, because it is the best at grouping by performance and the second best at predicting student errors with $support=0$.

To sum up, as explained above, the best clustering method (events by zone) can manage prediction errors around 0.3 (with $support=0$ and $confidence \approx 0.5$) at the beginning of a learning session and even better later.  This is because as the learning session progresses, the model will tend to fit the students under supervision better. Hence, we can claim that the first goal posed at the beginning of section 5 has been reached, i.e., it was verified that the prediction error is acceptable for tutoring purposes. In addition, we can also claim that the second goal has also been reached, because events by zone clustering classifies students in groups that will require a different tutoring feedback. 

\section{Conclusions}\label{sec:conclusion}
This paper presents a model that can predict student actions in procedural training environments. Additionally, this paper explains how this model is integrated into an ITS architecture and how it can be used to improve the tutoring feedback by anticipating student errors as long as this is pedagogically convenient.

The collective student model is created from student logs by clustering logs and computing an extended automaton for each resulting cluster. We should highlight that there are few ITSs in the literature that rely on data mining techniques to enhance their tutoring feedback. 

The proposed model has been validated using the student logs collected in a 3D virtual laboratory for teaching biotechnology. As a result of this validation, we concluded that the model can provide reasonably good predictions and support tutoring feedback that is more adapted to each student type.

An application that displays the collective student model would be very useful for facilitating the definition of the tutoring strategy. In this way, the instructor could visualize when students make more mistakes or which part of the practical assignment students find easier. Based on this information, the instructor could decide where and what tutoring feedback the ITS should provide. Additionally, this could also help the instructor to improve his or her own teaching.

Another line of future work will be to validate an ITS built upon the proposed model in order to evaluate the tutoring feedback induced by the proposed model.

\ifCLASSOPTIONcompsoc
  \section*{Acknowledgments}
\else
  \section*{Acknowledgment}
\fi

Riofr\'io would like to acknowledge financial support from the Ecuadorian Secretariat of Higher Education, Science, Technology and Innovation (SENESCYT). Thanks to Rachel Elliot for her comments on the paper.

\ifCLASSOPTIONcaptionsoff
  \newpage
\fi

\begin{IEEEbiography}[{\includegraphics[width=1in,height=1.25in,clip,keepaspectratio]{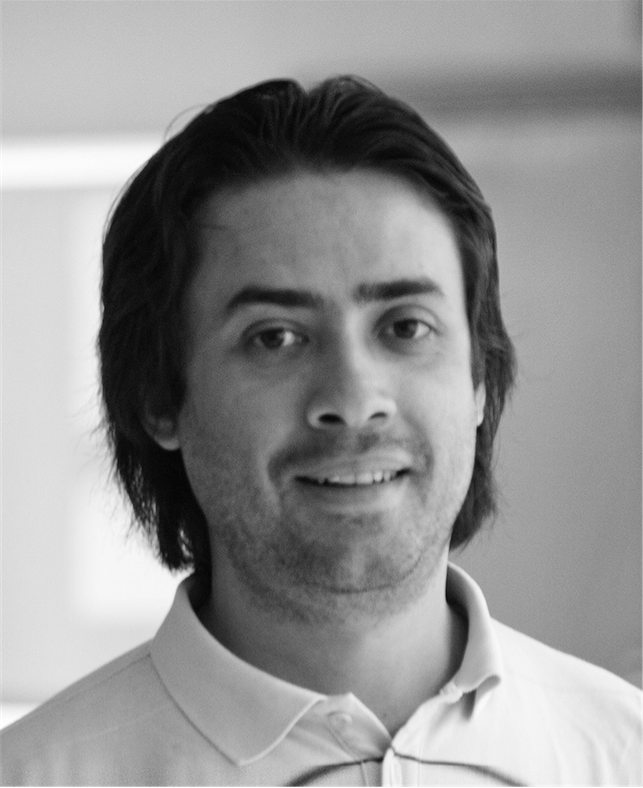}}]{Diego Riofr\'io}
is a PhD in Computer Science student at UPM. He holds a MSc in Computer Science (UPM 2012) and a BSc in Computer Engineering (Escuela Polit\'ecnica Nacional of Ecuador 2006). From 2004 to 2010 he was a R\&D engineer at private companies in Ecuador. His main research interests include Educational and Training Virtual Environments, Intelligent Tutoring Systems and Virtual Worlds.
\end{IEEEbiography}

\begin{IEEEbiography}[{\includegraphics[width=1in,height=1.25in,clip,keepaspectratio]{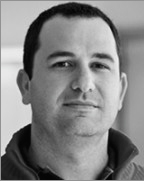}}]{Jaime Ramirez}
is associate professor at the School of Computer Science, UPM. He holds a MSc in Computer Science (UPM, 1996) and a PhD in Computer Science (UPM 2002). His main research interests include Ontology Engineering, and Adaptive Systems with an special focus on Intelligent Tutoring Systems combined with 3D Virtual Environments.
\end{IEEEbiography}

\begin{IEEEbiography}[{\includegraphics[width=1in,height=1.25in,clip,keepaspectratio]{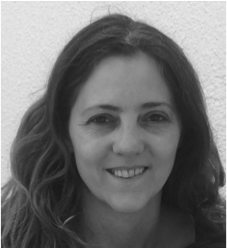}}]{Marta Berrocal} 
received a BSc in Biological Sciences from the Universidad Complutense de Madrid (UCM) in 1996, and a PhD from the Universidad Polit\'ecnica de Madrid (UPM) in 2002. Since 2007 she has been assistant professor of Biochemistry and Biotechnology at the UPM's School of Forestry Engineering and  project coordinator of the UPM Biotechnology Virtual Laboratory Creation and Development Educational Innovation Group.
\end{IEEEbiography}

\vfill



\begin{thebibliography}{1}
\bibitem{Romero2007}
C.~Romero and S.~Ventura, ``{Educational data mining: A survey from 1995 to
  2005},'' \emph{Expert Systems with Applications}, vol.~33, no.~1, pp.
  135--146, 2007.

\bibitem{Romero2010}
C.~Romero and S.~Ventura, ``{Educational Data Mining: A Review of the State of
  the Art},'' \emph{IEEE Transactions on Systems, Man, and Cybernetics, Part C
  (Applications and Reviews)}, vol.~40, no.~6, pp. 601--618, 2010.

\bibitem{Baker2014}
R.~S. Baker, ``{Educational Data Mining: An Advance for Intelligent Systems in
  Education},'' \emph{Intelligent Systems, IEEE}, vol.~29, no.~3, pp. 78--82,
  2014.
  
\bibitem{Vygotsky1978}
L.~S. Vygotsky, \emph{{Mind in society: The development of higher psychological
  processes}}.\hskip 1em plus 0.5em minus 0.4em\relax Harvard university press,
  1978.

\bibitem{Olneygift2014}
A.~M. Olney, ``{Scaffolding Made Visible},'' in \emph{Design Recommendations
  for Intelligent Tutoring Systems}.\hskip 1em plus 0.5em minus 0.4em\relax
  Orlando: U.S. Army Research Laboratory, 2014, ch.~26, pp. 327--340.

\bibitem{Holdengift2014}
H.~K. Holden and A.~M. Sinatra, ``{A Guide to Scaffolding and Guided
  Instructional Strategies for ITSs},'' in \emph{Design Recommendations for
  Intelligent Tutoring Systems}.\hskip 1em plus 0.5em minus 0.4em\relax
  Orlando: U.S. Army Research Laboratory, 2014, ch.~22, pp. 265--281.

\bibitem{Romero2010a}
C.~Romero, S.~Ventura, M.~Pechenizkiy, and R.~S. Baker, \emph{{Handbook of
  Educational Data Mining}}.\hskip 1em plus 0.5em minus 0.4em\relax CRC Press,
  2010.

\bibitem{Perera2009}
D.~Perera, J.~Kay, I.~Koprinska, K.~Yacef, and O.~R. Za{\&}{\#}x0EF;ane,
  ``{Clustering and Sequential Pattern Mining of Online Collaborative Learning
  Data},'' pp. 759--772, 2009.

\bibitem{Tang2005}
T.~Y. Tang and G.~McCalla, ``{Smart Recommendation for an Evolving E-Learning
  System: Architecture and Experiment},'' \emph{International Journal on
  ELearning}, vol.~4, no.~1, pp. 105--129, 2005.

\bibitem{Godoy2010}
D.~Godoy and A.~Amandi, ``{Link Recommendation in E-Learning Systems Based on
  Content-Based Student Profiles},'' in \emph{Handbook of Educational Data
  Mining}.\hskip 1em plus 0.5em minus 0.4em\relax CRC Press, 2010, ch.~19, pp.
  273--286.

\bibitem{Fournier-Viger2013}
P.~Fournier-Viger, R.~Nkambou, E.~M. Nguifo, A.~Mayers, and U.~Faghihi, ``{A
  Multiparadigm Intelligent Tutoring System for Robotic Arm Training},''
  \emph{Learning Technologies, IEEE Transactions on}, vol.~6, no.~4, pp.
  364--377, 2013.

\bibitem{Dekker2009}
G.~W. Dekker, M.~Pechenizkiy, and J.~M. Vleeshouwers, ``{Predicting Students
  Drop Out: A Case Study.}'' in \emph{International Working Group on
  Educational Data Mining}.\hskip 1em plus 0.5em minus 0.4em\relax ERIC, 2009.

\bibitem{Romero2013a}
C.~Romero, P.~G. Espejo, A.~Zafra, J.~R. Romero, and S.~Ventura, ``{Web usage
  mining for predicting final marks of students that use Moodle courses},''
  \emph{Computer Applications in Engineering Education}, vol.~21, no.~1, pp.
  135--146, mar 2013.

\bibitem{Lara2014}
J.~A. Lara, D.~Lizcano, M.~A. Mart{\'{i}}nez, J.~Pazos, and T.~Riera, ``{A
  system for knowledge discovery in e-learning environments within the European
  Higher Education Area – Application to student data from Open University of
  Madrid, UDIMA},'' \emph{Computers {\&} Education}, vol.~72, no.~0, pp.
  23--36, mar 2014.

\bibitem{Antunes2008}
C.~Antunes, ``{Acquiring Background Knowledge for Intelligent Tutoring
  Systems.}'' in \emph{EDM}, 2008, pp. 18--27.

\bibitem{Hershkovitz2009}
A.~Hershkovitz and R.~Nachmias, ``{Learning about online learning processes and
  students' motivation through Web usage mining},'' \emph{Interdisciplinary
  Journal of Knowledge and Learning Objects}, pp. 197--214, 2009.

\bibitem{Arroyo2009}
I.~Arroyo, D.~G. Cooper, W.~Burleson, B.~P. Woolf, K.~Muldner, and
  R.~Christopherson, ``{Emotion Sensors Go To School},'' in \emph{Proceedings
  of the 2009 Conference on Artificial Intelligence in Education: Building
  Learning Systems That Care: From Knowledge Representation to Affective
  Modelling}.\hskip 1em plus 0.5em minus 0.4em\relax Amsterdam, The
  Netherlands, The Netherlands: IOS Press, 2009, pp. 17--24.

\bibitem{Barnes2010}
T.~Barnes and J.~Stamper, ``{Automatic Hint Generation for Logic Proof Tutoring
  Using Historical Data},'' \emph{Journal of Educational Technology {\&}
  Society}, vol.~13, no.~1, pp. 3--12, 2010.

\bibitem{Mavrikis2010}
M.~Mavrikis, ``{Modelling student interactions in intelligent learning
  environments: constructing bayesian networks from data},''
  \emph{International Journal on Artificial Intelligence Tools}, vol.~19,
  no.~06, pp. 733--753, dec 2010.

\bibitem{Porayska-Pomsta2013}
K.~Porayska-Pomsta, M.~Mavrikis, S.~D'Mello, C.~Conati, and R.~S. J.~d. Baker,
  ``{Knowledge Elicitation Methods for Affect Modelling in Education},''
  \emph{Int. J. Artif. Intell. Ed.}, vol.~22, no.~3, pp. 107--140, 2013.

\bibitem{Bogarin2014}
A.~Bogar{\'{i}}n, C.~Romero, R.~Cerezo, and M.~S{\'{a}}nchez-Santill{\'{a}}n,
  ``{Clustering for Improving Educational Process Mining},'' in
  \emph{Proceedings of the Fourth International Conference on Learning
  Analytics And Knowledge}.\hskip 1em plus 0.5em minus 0.4em\relax New York,
  NY, USA: ACM, 2014, pp. 11--15.

\bibitem{Pena-Ayala2014}
A.~Pe{\~{n}}a-Ayala, ``{Educational data mining: A survey and a data
  mining-based analysis of recent works},'' \emph{Expert Systems with
  Applications}, vol.~41, no.~4, pp. 1432--1462, mar 2014.

\bibitem{Sukhija2015}
K.~Sukhija, M.~Jindal, and N.~Aggarwal, ``{The recent state of educational data
  mining: A survey and future visions},'' pp. 354--359, 2015.

\bibitem{Razzaq2005}
L.~Razzaq, M.~Feng, G.~Nuzzo-Jones, N.~T. Heffernan, K.~R. Koedinger,
  B.~Junker, S.~Ritter, A.~Knight, C.~Aniszczyk, and S.~Choksey, ``{The
  Assistment project: Blending assessment and assisting},'' \emph{Proceedings
  of the 12th Annual Conference on Artificial Intelligence in Education}, pp.
  555--562, 2005.

\bibitem{Ritter2007}
S.~Ritter, J.~Anderson, K.~Koedinger, and A.~Corbett,
  ``{Cognitive Tutor: Applied research in
  mathematics education},'' \emph{Psychonomic
  Bulletin {\&} Review}, vol.~14, no.~2, pp. 249--255, 2007.
  
\bibitem{Ritter2015}
S.~Ritter, R.~Carlson, M.~Sandbothe, and S.~E. Fancsali, ``{Carnegie Learning's
  Adaptive Learning Products},'' in \emph{Educational Data Mining 2015: 8th
  International Conference on Educational Data Mining}, Madrid, 2015.

\bibitem{vicente2015}
F.~Vicente, S.~Adjei, T.~Colombo, and N.~Heffernan, ``{Building Models to
  Predict Hint-or-Attempt Actions of Students},'' in \emph{Educational Data
  Mining 2015: 8th International Conference on Educational Data Mining},
  Madrid, 2015.

\bibitem{Baker2012}
R.~S. Baker, S.~M. Gowda, M.~Wixon, J.~Kalka, A.~Z. Wagner, A.~Salvi,
  V.~Aleven, G.~W. Kusbit, J.~Ocumpaugh, and L.~Rossi, ``{Towards Sensor-Free
  Affect Detection in Cognitive Tutor Algebra.}'' \emph{International
  Educational Data Mining Society}, 2012.

\bibitem{Romero2013}
C.~Romero and S.~Ventura, ``{Data mining in education},'' \emph{Wiley
  Interdisciplinary Reviews: Data Mining and Knowledge Discovery}, vol.~3,
  no.~1, pp. 12--27, 2013.

\bibitem{DeAntonio2005}
A.~de~Antonio, J.~Ram{\'{i}}rez, R.~Imbert, and G.~M{\'{e}}ndez, ``{Intelligent
  virtual environments for training: An agent-based approach},''
  \emph{Multi-Agent Systems and Applications IV}, pp. 82--91, 2005.

\bibitem{Imbert2007}
R.~Imbert, L.~S{\'{a}}nchez, A.~de~Antonio, G.~M{\'{e}}ndez, and
  J.~Ram{\'{i}}rez, ``{A multiagent extension for virtual reality based
  intelligent tutoring systems},'' in \emph{International Conference on
  Advanced Learning Technologies}.\hskip 1em plus 0.5em minus 0.4em\relax IEEE,
  2007, pp. 82--84.

\bibitem{Clemente2011}
J.~Clemente, J.~Ram{\'{i}}rez, and A.~de~Antonio, ``{A proposal for student
  modeling based on ontologies and diagnosis rules},'' \emph{Expert Systems
  with Applications}, vol.~38, no.~7, pp. 8066--8078, jul 2011.

\bibitem{Clemente2011a}
J.~Clemente, ``{Una Propuesta de Modelado del Estudiante Basada en
  Ontolog{\'{i}}as y Diagn{\'{o}}stico Pedag{\'{o}}gico-Cognitivo no
  Mon{\'{o}}tono},'' Ph.D. dissertation, Universidad Polit{\'{e}}cnica de
  Madrid, 2011.

\bibitem{Clemente2014}
J.~Clemente, J.~Ram{\'{i}}rez, and A.~de~Antonio, ``{Applying a student
  modeling with non-monotonic diagnosis to Intelligent Virtual Environment for
  Training/Instruction},'' \emph{Expert Systems with Applications}, vol.~41,
  no.~2, pp. 508--520, feb 2014.

\bibitem{Rico2012}
M.~Rico, J.~Ram{\'{i}}rez, D.~Riofr{\'{i}}o, M.~Berrocal-Lobo, and A.~{De
  Antonio}, ``{An architecture for virtual labs in engineering education},'' in
  \emph{Global Engineering Education Conference (EDUCON), 2012 IEEE}, 2012, pp.
  1--5.

\bibitem{Mostow2006}
J.~Mostow and J.~Beck, ``{Some Useful Tactics to Modify, Map and Mine Data from
  Intelligent Tutors},'' \emph{Nat. Lang. Eng.}, vol.~12, no.~2, pp. 195--208,
  2006.

\bibitem{Mannila2000}
H.~Mannila and C.~Meek, ``{Global Partial Orders from Sequential Data},'' in
  \emph{Proceedings of the Sixth ACM SIGKDD International Conference on
  Knowledge Discovery and Data Mining}, KDD '00, New York, NY, USA: ACM, 2000, pp. 161--168.

\bibitem{Mannila1997}
H.~Mannila, H.~Toivonen, and A.~{Inkeri Verkamo}, ``{Discovery of Frequent
  Episodes in Event Sequences},'' \emph{Data Mining and Knowledge Discovery},
  vol.~1, no.~3, pp. 259--289, 1997.

\bibitem{Buijs2014}
J.~C. Buijs, B.~F. van Dongen, W.~M. van~der Aalst, and P, ``{Quality
  Dimensions in Process Discovery: The Importance of Fitness, Precision,
  Generalization and Simplicity},'' \emph{International Journal of Cooperative
  Information Systems}, vol.~23, no.~01, p. 1440001, 2014.
  
\bibitem{Rokach2005}
L.~Rokach, and O.~Maimon, ``{Clustering Methods},'' in \emph{Data Mining and Knowledge Discovery Handbook}, Boston, MA: Springer US, pp. 321–352, 2005

\bibitem{Han2012Ch10}
J.~Han, M.~Kamber, and J.~Pei, ``{Cluster Analysis: Basic Concepts and Methods},'' in \emph{Data mining: concepts and techniques}, Elsevier Inc, 3rd ed., pp. 443–494, 2012.

\bibitem{Han2012Ch11}
J.~Han, M.~Kamber, and J.~Pei, ``{Advanced Cluster Analysis},'' in \emph{Data mining: concepts and techniques}, Elsevier Inc, 3rd ed., pp. 497–542, 2012.

\bibitem{Tan2013Ch8}
P.-N.~Tan, M.~Steinbach, and V.~Kumar, ``{Data mining cluster analysis: Basic concepts and algorithms},'' in \emph{Introduction to data mining}, London: Pearson Education Limited, 1st ed., 2013.

\bibitem{Hall2009}
M.~Hall, E.~Frank, G.~Holmes, B.~Pfahringer, P.~Reutemann, and I.~H. Witten,
  ``{The WEKA Data Mining Software: An Update},'' \emph{SIGKDD Explor. Newsl.},
  vol.~11, no.~1, pp. 10--18, 2009. 

\bibitem{Microsoft}
Microsoft, ``{Microsoft Sequence Clustering Algorithm Technical Reference},''
  2016. [Online]. Available:
  \url{https://msdn.microsoft.com/en-us/library/cc645866.aspx}

\bibitem{Slonim:2002}
N.~Slonim, N.~Friedman, and N.~Tishby, ``{Unsupervised Document Classification
  Using Sequential Information Maximization},'' in \emph{Proceedings of the
  25th Annual International ACM SIGIR Conference on Research and Development in
  Information Retrieval}, ser. SIGIR '02.\hskip 1em plus 0.5em minus
  0.4em\relax New York, NY, USA: ACM, 2002, pp. 129--136.
\end{thebibliography}
\end{document}